\documentclass[nofootinbib,pra,twocolumn,showpacs,superscriptaddress,notitlepage,amsmath,amssymb]{revtex4-1}

\usepackage[colorlinks=true,citecolor=blue,linkcolor=blue,urlcolor=blue]{hyperref}
\usepackage{bbm,tikz}
\usepackage{multirow}

\usepackage{amsmath,amssymb,bm,mathtools}

\usepackage{tikz}

\def\SPT{\text{SPT}}

\def\bra#1{\mathinner{\langle{#1}|}}
\def\ket#1{\mathinner{|{#1}\rangle}}
\def\bs#1{\boldsymbol{#1}}
\def\inner#1{\mathinner{\langle{#1}\rangle}}
\def\ZZ{\mathbb Z}

\makeatletter 
    
\renewcommand\onecolumngrid{
\do@columngrid{one}{\@ne}%
\def\set@footnotewidth{\onecolumngrid}
\def\footnoterule{\kern-6pt\hrule width 1.5in\kern6pt}%
}

\makeatother 

\begin{document}

\title{Shortest Route to Non-Abelian Topological Order on a Quantum Processor}

\author{Nathanan Tantivasadakarn}
\thanks{Both authors contributed equally}
\affiliation{Walter Burke Institute for Theoretical Physics and Department of Physics, California Institute of Technology, Pasadena, CA 91125, USA}
\affiliation{Department of Physics, Harvard University, Cambridge, MA 02138, USA}

\author{Ruben Verresen}
\thanks{Both authors contributed equally}
\affiliation{Department of Physics, Harvard University, Cambridge, MA 02138, USA}

\author{Ashvin Vishwanath}
\affiliation{Department of Physics, Harvard University, Cambridge, MA 02138, USA}

\date{\today}

\begin{abstract}
A highly coveted goal is to realize emergent non-Abelian gauge theories and their anyonic excitations, which encode decoherence-free quantum information. While measurements in quantum devices provide new hope for scalably preparing such long-range entangled states, existing protocols using the experimentally established ingredients of a finite-depth circuit and a single round of measurement produce only Abelian states. Surprisingly, we show there exists a broad family of non-Abelian states---namely those with a Lagrangian subgroup---which can be created using these same minimal ingredients, bypassing the need for new resources such as feed-forward. To illustrate that this provides realistic protocols, we show how $D_4$ non-Abelian topological order can be realized, e.g., on Google's quantum processors using a depth-11 circuit and a single layer of measurements. Our work opens the way towards the realization and manipulation of non-Abelian topological orders, and highlights counter-intuitive features of the complexity of non-Abelian phases.
\end{abstract}

\maketitle

\begin{figure}
    \centering
    \includegraphics[scale=0.55]{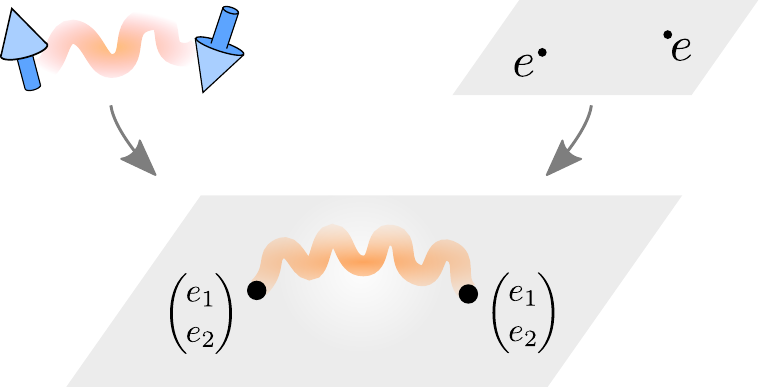}
    \caption{\textbf{Artist's impression of a non-Abelion.} Non-Abelian anyons as in $D_4$ topological order bring together two ingredients in a remarkable mix: Bell pairs and gauge charges. Non-Abelions transform under a non-trivial matrix representation of the gauge group, leading to a topological degeneracy. 
    The Bell pair is a robust consequence of forming a gauge neutral singlet.
    In this work we show how to efficiently prepare $D_4$ non-Abelian order with a single layer of measurement, whereby non-Abelion entanglement serves as a smoking gun.
    }
    \label{fig:cartoon}
\end{figure}

The quantum statistics of particles is a foundational concept with far-reaching ramifications, and in two spatial dimensions, a remarkably rich set of `anyonic statistics' arises \cite{Leinaas77,Wilczek82}. Although not realized by fundamental particles, anyons emerge as effective quasi-particles in two-dimensional condensed matter systems, most notably the fractional Quantum Hall effect \cite{Halperin20}. The most exotic extension of quantum statistics occurs with \emph{non-Abelian} anyons \cite{Bais80,moore_classical_1989,Witten89,Kitaev_2003}
which always come in degenerate quantum states (Fig.~\ref{fig:cartoon}). Consequently, while braiding Abelian anyons only lead to a phase factor, braiding `non-Abelions' leads to a matrix action on the degenerate states. This has evoked dreams of a physically fault-tolerant route to perform quantum computing, with quantum gates being executed by the motion of non-Abelian anyons \cite{Nayak_RMP}. However, a key obstacle is finding states of matter hosting such non-Abelions, called \emph{non-Abelian topological order} \cite{Wenbook}. The most compelling candidates so far are certain fractional quantum Hall states in the second Landau level ($\nu=5/2,\,12/5$) \cite{MooreRead91,Halperin20,Nayak_RMP}. However, non-Abelian states are more fragile compared compared to their Abelian counterparts \cite{Nayak_RMP,Dean08} and the extreme conditions required to create quantum Hall states, combining high magnetic fields, pristine samples and millikelvin temperatures, all call for new approaches to creating such quantum states.

Meanwhile, the significant advances in near-term quantum devices \cite{NISQ} open up the possibility of realizing non-Abelian states, not from cooling into the ground states, but from controlled quantum gates that entangle a product state to resemble ground states with non-Abelian excitations. Indeed, recent theory and experimental work has shown how certain Abelian states can be created in this way, in particular the toric code topological order \cite{Verresen21prediction,Semeghini21,Satzinger21}. However, the general strategy adopted in these works is essentially a form of adiabatic state preparation whose depth is required to scale with system size \cite{BravyiHastingsVerstraete06}, a formidable requirement when one wants to scale system size with limited depth quantum circuits.

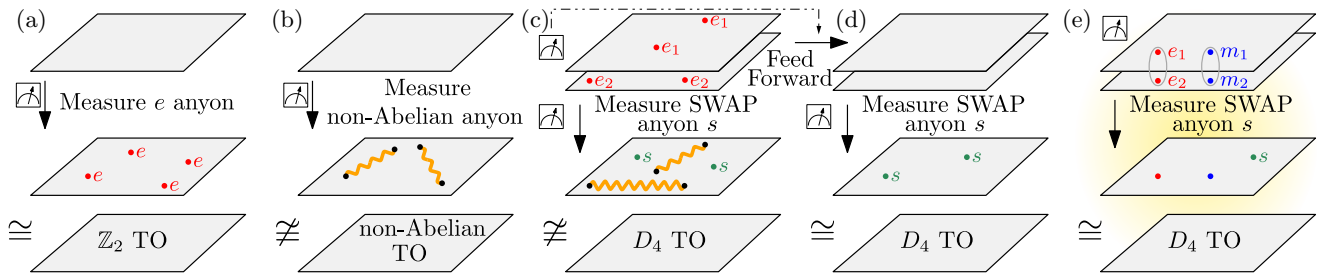
\begin{figure*}[t!]
    \centering
    \begin{tikzpicture}
    \node at (0,0) 
    {\includegraphics[scale=0.9]{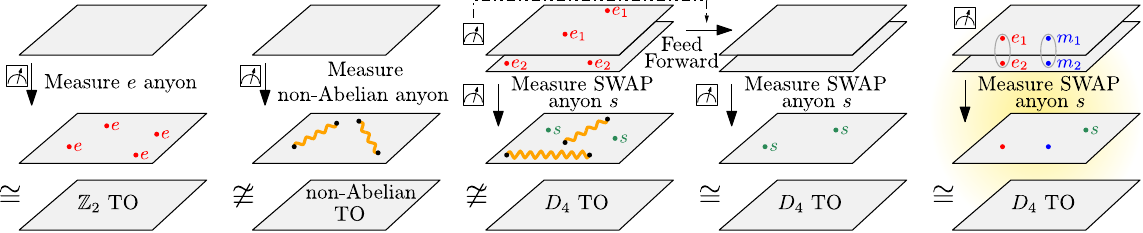}};
    \node at (-8.4,1.6) {(a)};
    \node at (-5,1.6) {(b)};
    \node at (-1.7,1.6) {(c)};
    \node at (2.5,1.6) {(d)};
    \node at (5.5,1.6) {(e)};
    \end{tikzpicture}
    \caption{\textbf{Topological order from measurement.} (a) By measuring the star term $A_v$ \eqref{eq:AvBp} of the toric code, $\mathbb Z_2$ topological order (TO) is obtained regardless of the measurement outcome. A clean toric code is achieved by pairing up $e$-anyons by a feed-forward of single-site Pauli operators. (b) In contrast, measuring non-Abelian gauge charges gives rise to topological degeneracies and non-Abelion entanglement (Fig.~\ref{fig:cartoon}). Removing these requires a unitary circuit whose depth scales with system size \cite{Shi19}. (c) One route to non-Abelian topological order is to first prepare two copies of the toric code by measuring the $e$-anyons in each layer. We obtain non-Abelian $D_4 \cong \left( \mathbb Z_2 \times \mathbb Z_2 \right) \rtimes \mathbb Z_2$ topological order if we gauge the swap symmetry of the two layers. However, the Abelian anyon defects of the bilayer then become non-Abelian defects. (d) The aforementioned protocol does work if we use feed-forward to obtain two \emph{clean} toric codes before gauging the swap symmetry.
    (e) In this work, we point out that one can obtain the same phase of matter \emph{without} feed-forward, using only a \emph{single} measurement layer. The key is to first prepare the two copies of the toric code by measuring the appropriate charges and fluxes which are \emph{invariant} under the swap symmetry; subsequently gauging the swap symmetry does thus not introduce non-Abelian anyons. In fact, we can measure all three anyons at once. See Fig.~\ref{fig:D4} or Eq.~\eqref{eq:D4} for an explicit and realistic circuit.}
    \label{fig:TC}
\end{figure*}

Remarkably, a workaround exists which allows to create certain topological orders in a time independent of system size. For instance, the aforementioned toric code is obtained at once by simply measuring its two commuting stabilizers on the links of the square lattice \cite{Gottesman97,Kitaev_2003,Raussendorf05,Aguado08,Piroli21}:
\begin{equation}
A_v =
\raisebox{-20pt}{
\begin{tikzpicture}
\draw[-,black!10,line width=1.2] (-0.8,0) -- (0.8,0);
\draw[-,black!10,line width=1.2] (0,-0.8) -- (0,0.8);
\node at (-0.4,0.05) {$\sigma^z$};
\node at (0.5,0.05) {$\sigma^z$};
\node at (0.1,0.55) {$\sigma^z$};
\node at (0.1,-0.45) {$\sigma^z$};
\end{tikzpicture}
} \quad \textrm{and}
\quad
B_p = \raisebox{-20pt}{
\begin{tikzpicture}
\draw[-,black!10,line width=1.2] (-0.5,-0.5) -- (0.5,-0.5) -- (0.5,0.5) -- (-0.5,0.5) -- (-0.5,-0.5);
\node at (-0.4,0.05) {$\sigma^x$};
\node at (0.6,0.05) {$\sigma^x$};
\node at (0.1,0.53) {$\sigma^x$};
\node at (0.1,-0.45) {$\sigma^x$};
\end{tikzpicture}
}
.
\label{eq:AvBp}
\end{equation}
Stronger yet, starting from a product state $\ket{\psi} = \ket{+}^{\otimes N}$, one needs to measure only $A_v$ (see Fig.~\ref{fig:TC}a). The random measurement outcomes for $A_v$ do not affect the $\mathbb Z_2$ topological order: the resulting `$e$-anyons' ($A_v=-1$) are static Abelian charges which simply redefine our notion of vacuum state. If one moreover wants to prepare the `clean' case ($A_v=B_p=1$), we note that these $e$-anyons come in pairs and can be removed by a single feed-forward layer of $\sigma^x$-string operators \cite{Kitaev_2003}.

The above approach generalizes to various other Abelian topological orders \cite{Bolt16}. However, the richer non-Abelian topological order does not admit such a simple stabilizer description, but at best only a commuting projector Hamiltonian \cite{Kitaev_2003,levin_string-net_2005}. Indeed, due to the intrinsic degeneracies associated to non-Abelions, the excited states do not resemble the ground state---in fact, they are not the ground state of any local gapped Hamiltonian. Hence, if one naively measures the terms in their parent Hamiltonian, one typically produces non-Abelian charges (Fig.~\ref{fig:TC}b), which cannot even be paired up by any finite-depth unitary string operator \cite{Shi19}. Intuitively, this is linked to the `Bell pair' mentioned in Fig.~\ref{fig:cartoon}.

This raises the question: is non-Abelian topological order out of the reach of a simple measurement protocol? Partial results are known where measurement helps: it has recently been shown that certain non-Abelian topological orders can be obtained in finite time by \emph{several} layers of measurement, interspersed with feed-forward \cite{Rydberg,measureSPT,Bravyi22,Lu2022}. In light of these sophisticated protocols, and the aforementioned issue, it seems nigh impossible to obtain non-Abelian topological order from a \emph{single} layer of measurements. This is of more than mere conceptual interest: feed-forward remains a very costly ingredient, with many quantum simulators and computing platforms not yet allowing for it. A protocol which avoids it, as for the toric code above, is thus of conceptual and practical significance.

Here, we show that a class of non-Abelian topological order can be created by a single layer of measurements, thereby thus not requiring feed-forward. Surprisingly, this shows that there exists a class of non-Abelian states which are no more complex to prepare than their Abelian counterparts, but nevertheless display richer behavior. 

As a conceptually simple route towards non-Abelian order, let us imagine starting with \emph{two} copies of the toric code. These can be prepared by measuring the star term $A_v$ \eqref{eq:AvBp} on each layer, producing $e_1$- and $e_2$-anyons on the two layers (Fig.~\ref{fig:TC}c). Such a bilayer has a natural `swap' symmetry interchanging the two copies. If this \emph{global physical} symmetry were turned into a \emph{local gauge} symmetry, we would achieve non-Abelian topological order. Indeed, the $e_1$ and $e_2$ anyons then transform as a doublet under the gauge group, which can be identified with $D_4 = \left( \mathbb Z_2 \times \mathbb Z_2 \right) \rtimes \mathbb Z_2$ \cite{BarkeshliBondersonChengWang2019,SM}. To obtain this gauge symmetry, we can proceed as in the toric code case, i.e., by simply measuring the gauge charge operator (or more precisely, its Gauss law operator); soon we make this more explicit. This has two effects: first, this produces a speckle of Abelian anyons associated to the swap gauge symmetry; this is as harmless as in the toric code case. A more serious issue is that the Abelian anyons of the toric code now turn into non-Abelian anyon defects (Fig.~\ref{fig:TC}c).

\begin{figure*}
\begin{tikzpicture}
\node at (0,0) {\includegraphics[scale=0.40]{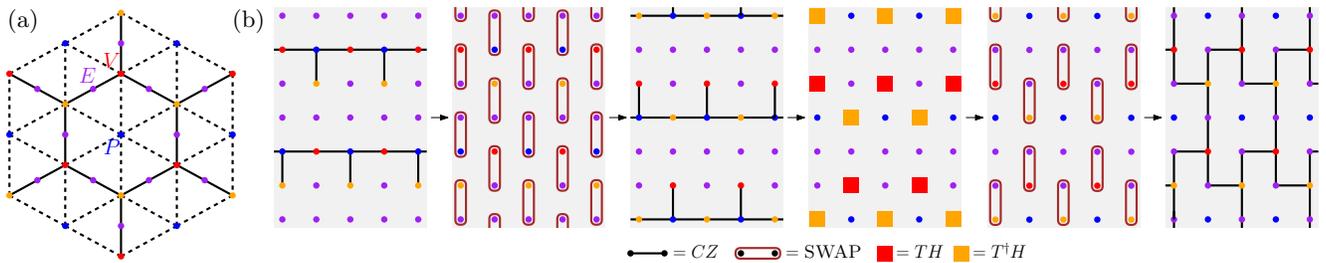}};
\node at (-8.5,1.5) {(a)};
\node at (-5.5,1.5) {(b)};
\end{tikzpicture}
\caption{\textbf{Preparing non-Abelian $D_4$ topological order with a single layer of measurement: from theory to Sycamore chip.} (a) The intuitive approach sketched in Fig.~\ref{fig:TC}e is formalized in Eq.~\eqref{eq:D4} for qubits on edges ($E$), vertices ($V$) and plaquettes ($P$) of the honeycomb lattice. Initializing all qubits in $\ket{+}$, the circuit consists of three steps: 1) $CZ$ gates connecting plaquettes to vertices (dashed lines), which form the dice lattice. 2) $TH$ on the red vertices, and $T^\dagger H$ on the orange vertices (to wit, $T \propto e^{-i\frac{\pi}{8}Z}$). 3) $CZ$ gates connecting vertices to edges (solid lines), which forms the heavy-hex lattice. Finally, we measure all vertices and plaquettes in the $X$-basis, producing $D_4$ topological order for any measurement outcome, using only 9 layers of non-overlapping two-body gates. (b) Implementation on device with square lattice connectivity, e.g., Google's Bristlecone and Sycamore chips. The first three panels prepare the dice lattice cluster state, the fourth panel performs the single-site basis rotation, and the last two panels applies the cluster state entangler for the heavy-hex lattice. Finally, measuring all but the purple-color qubits produces $D_4$ non-Abelian topological order. This protocol is independent of system size, requiring only 11 layers of non-overlapping two-body gates.}
\label{fig:D4}
\end{figure*}

So far, the above example thus hits on the same stumbling block: in the quest to produce non-Abelian order via measurement, we produce defects which destroy the desired phase of matter. One possible solution is to clean up the $e$-anyon defects before gauging the swap symmetry (Fig.~\ref{fig:TC}d); this gives a multi-step measurement protocol with feed-forward \cite{Rydberg,measureSPT,Bravyi22,Lu2022} which---while interesting---we here seek to avoid. We surmise that this stumbling block cannot be avoided if one measures only charges. However, we show the issue can be resolved by using the larger freedom of measuring charges \emph{or} fluxes (to wit, the fluxes of the toric code are also called $m$-anyons, as detected by $B_p$ in Eq.~\eqref{eq:AvBp}). Indeed, rather than producing the toric code bilayer by measuring $e_1$ and $e_2$, we can also produce it by measuring a different set of Abelian anyons: the composites $e_1e_2$ and $m_1m_2$ (Fig.~\ref{fig:TC}e). Crucially, these are a singlet under the swap symmetry. Hence, now proceeding as before, measuring the `swap anyons' produces \emph{only} Abelian defects. We have thereby produced $D_4$ topological order in finite time, without feed-forward! Observe that this approach works even if we measure the anyons $\{e_1e_2,m_1m_2,s\}$ all at once.

Let us now turn the above conceptual discussion into a concrete protocol for preparing $D_4$ topological order for qubits living on the edges ($E$) of the honeycomb lattice. To effectively measure the type of many-body operators discussed above, we will use two-body entangling gates and perform single-site measurements on ancilla qubits on the vertices ($V$) and plaquettes ($P$) of the honeycomb lattice. We claim that the topological order is obtained by the following sequence (Fig.~\ref{fig:D4}a):
\begin{equation}
\ket{D_4}_E = \bra{x}_{PV}\prod_{\langle v,e\rangle}CZ_{ve} \prod_v e^{\pm \frac{\pi i}{8} Z_v} H_v  \; \prod_{\langle p,v\rangle} CZ_{pv} \ket{+}_{PEV}, \label{eq:D4}
\end{equation}
where $X,Y,Z$ denote Pauli matrices, $H$ is the Hadamard gate, $CZ$ is the Controlled-$Z$ gate, and $x = \pm 1$ denotes the arbitrary outcome upon measuring all the ancillas in the $X$-basis.

We can break the above procedure down into three steps. First, performing $CZ_{pv}$ prepares the dice lattice cluster state whereby measuring the plaquettes results in the color code. This is unitarily equivalent to two copies of the toric code \cite{Kubica15}, playing the role of the bilayer in Fig.~\ref{fig:TC}e. The single-site gates on the vertices rotate the color code into a basis where the ``swap" symmetry is realized by $\prod_{v} X_v$.
Lastly, we gauge this symmetry by measuring its associated Gauss law operator on each vertex, $X_v \prod_{e \supset v} Z_e$, which is achieved by a single-site measurement preceded by the $CZ_{ve}$ unitary.

Importantly, any measurement in Eq.~\eqref{eq:D4} can be delayed to the last step. A similar formula appeared in Ref.~\onlinecite{Rydberg}, with the crucial difference that the single-site rotation was different. As a consequence, the latter requires feed-forward, corresponding to the scenario in Fig.\ref{fig:TC}d.

Certain quantum processors have restricted connectivity, and might thus not be able to directly apply the gates in Fig.\ref{fig:D4}a. In such cases it is still possible to create the $D_4$ state by using SWAP gates to attain the desired connectivity. To illustrate this, we propose an implementation for Google's quantum processor, which has the connectivity of a square lattice as shown in Fig.\ref{fig:D4}b. We find that the non-Abelian state can be prepared with a two-body depth of 11 layers, independent of total chip size. (This becomes 13 layers once we decompose the SWAP layers into Google's native $CZ$ gates; see the Supplemental Materials \cite{SM} for further discussion.)

While we have discussed the minimal case of $D_4$ topological order in great detail, we note that the idea of our efficient protocol extends to other topological orders which admit a so-called \emph{Lagrangian subgroup} \cite{KapustinSaulina11,Levin13}. This is defined to be a subgroup of Abelian anyons with trivial self and mutual statistics such that every other anyon in the theory braids non-trivially with it. In the case of $D_4$, this corresponds to the group generated by $\{e_1e_2,m_1m_2,s\}$ as encountered in Fig.~\ref{fig:TC}e. Phrased in the language of quantum doubles \cite{Kitaev_2003}, $e_1e_2$ and $s$ correspond to the sign representations of $D_4$, while $m_1m_2$ corresponds to the conjugacy class of the center of $D_4$ \cite{BarkeshliBondersonChengWang2019,SM}. It is known that if one condenses the anyons in the Lagrangian subgroup, one obtains a trivial state. By playing this argument in reverse, one can argue that measuring the Gauss law operators associated to these anyons, one obtains its non-Abelian topological order with only a single layer of measurement \cite{SM}. Other examples which can in principle be obtained in this way are, say, the quaternion $Q_8$ quantum double\cite{SM}, or even the doubled Ising topological order \cite{kitaev_anyons_2005,levin_string-net_2005} (by measuring the Gauss law for $\epsilon$ and $\bar \epsilon$, though this requires physical fermions). It would be interesting to work out explicit protocols amenable to quantum processors, as we did for $D_4$ above.

In conclusion, we have established the simplest route to non-Abelian topological order. While the preparation time for a purely unitary protocol must scale with system size, we found that the minimal non-unitary element of a single measurement layer could efficiently prepare certain non-Abelian orders. This furthermore avoids the need for feed-forward which is intrinsic to multi-measurement approaches \cite{Rydberg,measureSPT,Bravyi22,Lu2022}. For the illustrative case of $D_4$, we found that roughly ten unitary layers (prior to single-site measurements) were already sufficient, even for realistic qubit connectivity as on the Google chips. Naturally, it would be worthwhile to work out concrete protocols for other existing architectures. On the  conceptual side, the existence of a single-shot protocol for certain non-Abelian states motivates us to identify the minimal number of measurement layers (alongside finite-depth unitaries) for obtaining various types of quantum states. We will examine this measurement-induced hierarchy of quantum states in a forthcoming work \cite{hierarchy}.

Lastly, if a non-Abelian state is realized, how do we tell? One interesting probe is the aforementioned non-Abelion entanglement (Fig.~\ref{fig:cartoon}), which we can now turn into an advantage. Indeed, the successful preparation of non-Abelian order can be confirmed by noting that if we insert non-Abelian excitations, the entanglement entropy is changed according to its quantum dimension \cite{Kitaev06b}. For instance, for our particular $D_4$ protocol (Eq.~\eqref{eq:D4}), this is achieved by acting with Pauli-$Z$ operators on the vertices at any point prior to the single-site rotations. That such a deceptively simple tweak can have such a drastic consequence underlines the exotic nature of non-Abelian states, and points the way to the first realization and detection in a quantum simulator.

\vspace{5pt}

\begin{acknowledgments}
The authors thank Ryan Thorngren for collaboration on a related work \cite{measureSPT}. NT is supported by the Walter Burke Institute for Theoretical Physics at Caltech. RV is supported by the Harvard Quantum Initiative Postdoctoral Fellowship in Science and Engineering, and RV and AV by the Simons Collaboration on Ultra-Quantum Matter, which is a grant from the Simons Foundation (651440, AV).
\end{acknowledgments}

\bibliography{bib.bib}

\appendix
\onecolumngrid

\newpage

\section{Group theory of $D_4$}

\begin{figure}[h!]
    \centering
    \includegraphics{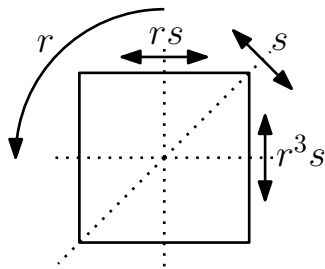}
    \caption{$D_4$ realizes the symmetries of the square. As a semidirect product $(\ZZ_2 \times \ZZ_2) \rtimes \ZZ_2$, the vertical and horizontal reflections ($rs$ and $r^3s$) are exchanged under the diagonal reflection $s$.}
    \label{fig:D4square}
\end{figure}

The group $D_4$ can be defined abstractly as generated by elements $r$ and $s$ which satisfy $r^4 = s^2 = (sr)^2 =1$. As symmetries of the square, $r$ rotates the square by $90^{\circ}$ degrees and $s$ performs a diagonal reflection, as shown in Fig.~\ref{fig:D4square}. The group $D_4$ can also be seen as a semidirect product $(\ZZ_2 \times \ZZ_2) \rtimes \ZZ_2$, where the vertical and horizontal reflections $rs$ and $r^3s$ are swapped under diagonal symmetry $s$.

The group admits five irreducible representations (irreps). Other than the trivial irrep $\bs 1$ and the faithful two-dimensional irrep $\bs 2$ where 
\begin{align}
        r& = \begin{pmatrix} i & 0\\0&-i \end{pmatrix} \qquad \textrm{and} \qquad s = \begin{pmatrix} 0 & 1\\1&0 \end{pmatrix},
\end{align}
there are three sign representations, which we will label $\bs s_1$, $\bs s_2$ and $\bs s_3$. Each sign rep is uniquely defined by its ``kernel", the subgroup on which the sign rep acts trivially.
\begin{enumerate}
\item $\bs s_1$ has kernel $\{1,r,r^2,r^3\}$ meaning it is represented by $s=rs=r^2=r^3s=-1$,
\item $\bs s_2$ has kernel $\{1,r^2,s,r^2s\}$ meaning it is represented by $r=r^3=rs=r^3s=-1$,
\item $\bs s_3$ has kernel $\{1,r^2,rs,r^3s\}$ meaning it is represented by $r=r^3 =s=r^2s=-1$.
\end{enumerate}

The group also admits five conjugacy classes, $[1]=\{1\}$, $[r^2]=\{r^2\}$, $[r]=\{r,\bar r\}$, $[s]=\{s,\bar r^2s\}$ $[rs]=\{rs,r^3s\}$.

\section{Correspondence between anyons of bilayer Toric Code and anyons of $D_4$ TO}

Mathematically, the anyons in the $G$ topological order (TO) corresponds to irreducible representations of the quantum double $\mathcal D(G)$. Each anyon can be given two labels: a conjugacy class $[g]$ and an irreducible representation of its centralizer $\pi_g$. A pure charge corresponds to the trivial conjugacy class with a choice of an irrep of $G$, while a pure flux corresponds a trivial irrep and a choice of conjugacy class. The quantum dimension of the anyon is given by the size of the conjugacy class times the dimension of the irrep. For $G=D_4$ enumerating all the possible choice of conjugacy classes and irreps gives a total of 22 anyons.

We are particularly interested in abelian anyons of $\mathcal D(D_4)$ and how they are related to the anyons that we measure in the toric code bilayer construction: $e_1e_2$, $m_1m_2$, and $s$. A complete treatment of how anyons map under gauging (which is implemented in this particular instance by measurement) can be found in Ref.~\onlinecite{BarkeshliBondersonChengWang2019}.

 First, without loss of generality we take the swap symmetry to be represented by the group element $s$. As shown in Fig. \ref{fig:D4square} it exchanges $rs$ and $r^3s$, the vertical and horizontal reflections. Since $rs$ and $r^3s$ generate a $\ZZ_2\times \ZZ_2$ subgroup, which is the kernel of $\bs s_3$, we identify the gauge charge of the swap symmetry with $\bs s_3$ as a gauge charge of the $D_4$ quantum double.
 
 Next, we note that $e_1e_2$ is a gauge charge of the bilayer toric code. In particular, it should be an irrep of the group $\ZZ_2 \times \ZZ_2$. In $D_4$, this is the subgroup generated by $rs$ and $r^3s$. Now, since $e_1e_2$ is the charged under the gauge transformation of both symmetries, while neutral under the diagonal symmetry $r^2$, it must therefore correspond to a representation where $rs=r^3s=-1$ while $r^2=1$. Moreover, $e_1e_2$ is neutral under the swap symmetry, meaning $s=1$. Comparing to the irreps of $D_4$, we therefore see that this matches the irrep $\bs s_2$. By a similar argument, we find that $e_1e_2s$ corresponds to the irrep $\bs s_1$.
 
 Finally, since $m_1m_2$ is a gauge flux of the bilayer toric code, it corresponds to a conjugacy class of $\ZZ_2 \times \ZZ_2$. Since $m_1$ and $m_2$ are associated to group elements $rs$ and $r^3s$, their product is therefore $r^2$. Hence, $m_1m_2$ corresponds to the $r^2$ conjugacy class of $D_4$.
 
 To conclude, the anyons we measure, $e_1e_2,m_1m_2,s$, generate eight anyons: $\{1,e_1e_2,m_1m_2,f_1f_2,s,e_1e_2s,m_1m_2s,f_1f_2s\}$. It is apparent that these anyons are all bosons and have trivial mutual braiding. Therefore, after gauging they are identified with eight abelian anyons of $D_4$ that forms a $\ZZ_2^3$ Lagrangian subgroup. The exact correspondence is summarized in Table \ref{tab:anyons}.
 
 It is also worth pointing out how non-Abelian anyons are generated in this correspondence. First consider the anyon $e_1$, which corresponds to the irrep $(-1,1)$ of $\ZZ_2 \times \ZZ_2$. Under the swap symmetry it transforms into $e_2$, corresponding to the irrep $(1,-1)$. Therefore, after gauging the swap symmetry, these two anyons combine into a single non-Abelian anyon with quantum dimension 2. This corresponds to the irrep $\bs 2$ of $D_4$. Note that in this case, the non-trivial action on the anyons means that it is not meaningful to attach the charge of the swap symmetry onto $[e_1]$. Moreover, this can be interpreted as the fusion rule $\bs 2 \times \bs s_3 = \bs 2$ for $D_4$ anyons.
 Similarly, the anyon $m_1$ and $m_2$ corresponds to the conjugacy class $\{ rs\}$ and  $\{r^3s\}$ respectively. After gauging, the conjugation of $s$ combines them into a single conjugacy class $[rs]$ of $D_4$, resulting in a non-Abelian gauge flux. 
 
 For further details on this specific correspondence, we refer to a thorough review in Sec. II of Ref.~\onlinecite{BulmashBarkeshli19}.

\begin{table}[]
    \centering
    \begin{tabular}{|c|c|c||c|c|c|c|}
        \hline
  \multicolumn{3}{|c||}{Bilayer TC with SWAP symmetry ($\ZZ_2^2 \rtimes \ZZ_2$)} & \multicolumn{3}{c|}{$D_4$ Quantum Double} & \multirow{2}{*}{dim} \\
  \cline{1-6}
  Orbit of anyon under SWAP & Stabilizer & SWAP charge &  Conj class & Centralizer & irrep&\\
  \hline
    $[1]$ & $\ZZ_2$ & 1          & $[1]$ & $D_4$ & $\bs 1$  &1\\
    $[e_1e_2]$ & $\ZZ_2$ & $s$      & $[1]$ & $D_4$ & $\bs s_1$ & 1\\
    $[e_1e_2]$ & $\ZZ_2$ &1        & $[1]$ & $D_4$ & $\bs s_2$ &1 \\
    $[1]$ & $\ZZ_2$ &$s$     & $[1]$ & $D_4$ & $\bs s_3$ & 1\\
    $[m_1m_2]$ & $\ZZ_2$ & 1           & $[{r^2}]$ & $D_4$ & $\bs 1$ &1 \\
    $[f_1f_2]$ & $\ZZ_2$ &$s$     & $[{r^2}]$ & $D_4$ & $\bs s_1$ &1 \\
    $[f_1f_2]$ & $\ZZ_2$ &1        & $[{r^2}]$ & $D_4$ & $\bs s_2$ & 1\\
    $[m_1m_2]$ & $\ZZ_2$ &$s$     & $[{r^2}]$ & $D_4$ & $\bs s_3$ &1 \\
    \hline
        $[e_1]=\{e_1,e_2\}$ & $\ZZ_1$ &1    & $[1]$ & $D_4$ & $\bs 2$  &2\\
        $[m_1]=\{m_1,m_2\}$ & $\ZZ_1$ & 1  & $[rs] = \{rs,r^3s \}$ & $\ZZ_2^2$ & $\bs 1$  &2\\
    \hline
    \end{tabular}
    \caption{Correspondence between anyons of Bilayer TC along with the SWAP symmetry charge, and anyons of $D_4$ TO. (Certain anyons are omitted for simplicity.)}
    \label{tab:anyons}
\end{table}

\section{Preparation of $D_4$ Topological order}

Here we prove that the protocol in the main text indeed prepares the $D_4$ quantum double with a single round of measurement. We first define the protocol on the vertices edges and faces of the triangular lattice, where the preparation is most natural.

We place qubits on the vertices, edges and plaquettes of the triangular lattice as in Fig.~3a of the main text. For convenience, the protocol is reproduced here:
\begin{align}
\ket{D_4}_E = \bra{x}_{PV}\prod_{\langle v,e\rangle}CZ_{ve} \prod_v e^{\pm \frac{\pi i}{8} Z_v} H_v  \; \prod_{\langle p,v\rangle} CZ_{pv} \ket{+}_{PEV}. \label{eq:D4app}
\end{align}
Namely, we start with a product state $\ket{+}$ for all qubits, apply the above quantum circuit, and perform projective measurements in the $x$-basis, where  $x=\pm$ labels the measurement outcomes on each vertex and plaquette. Here, the $\pm$ sign in $e^{\pm \frac{\pi i}{8} Z_v}$ denotes that the phase gate we perform takes an alternating sign depending on the vertex sublattice (colored red or orange in in Fig.~3a of the main text). 

The final state prepared, after a further Hadamard on all edges, is conveniently described as the simultaneous $+1$ eigenstate of the following ``stabilizers"\cite{Yoshida2016} defined for each vertex 

\begin{align}
    \bs A_p &= {\color{red}{x_p}} \times  \prod_{v \subset p}{\color{blue}x_v} \times \raisebox{-0.5\height}{\includegraphics[scale=0.9]{ApD4.pdf}}, 
    & \bs B^{(1)}_p &= \raisebox{-0.5\height}{\includegraphics[scale=0.9]{Bp1D4.pdf}},
    & \bs B^{(2)}_p &= \raisebox{-0.5\height}{\includegraphics[scale=0.9]{Bp2D4.pdf}}.
    \label{eq:D4stabilizers}
\end{align}
as we will momentarily derive. Note that without the $CZ$ operators in $\bs A_p$, these describe the stabilizers of three copies of the toric code. The $CZ$ operators couple the toric code in a non-trivial way that creates the $D_4$ TO (see Appendix \ref{app:D4TQD} for a further relation to the $\ZZ_2^3$ twisted quantum double and SPT phases).

In this model, although $\bs B_p^{(i)}$ commutes with all operators, two adjacent $\bs A_p$ operators only commute up to some product of $\bs B_{p'}^{(i)}$. For example, consider two adjacent plaquettes $p_L$ and $p_R$ sharing a vertical edge, then one has
\begin{align}
    \bs A_{p_L}\bs A_{p_R} = \bs B^{(1)}_{p_L}\bs B^{(2)}_{p_R} \bs A_{p_R}\bs A_{p_L}.
\end{align}
Nevertheless, one can still have a unique state which has eigenvalue $+1$ under all the above operators simultaneously, which is the state we prepare.

To facilitate in showing the above claim, we split the process into three steps
\begin{align}
   \ket{D_4}_E =  \bra{x}_V \prod_{\langle p,e\rangle}CZ_{pe} \ket{+}_E  \times e^{\pm \frac{\pi i}{8} Z_v} \prod_v H_v  \times \bra{x}_P\prod_{\langle v,p\rangle} CZ_{vp} \ket{+}_{PV}
\end{align}
The first step involves create a cluster state on the dice lattice by connecting each plaquette to each of the six vertices. Measuring all the plaquettes in the $x$-basis creates the color code. That is, the state
\begin{align}
    \ket{\text{CC}} = \bra{x}_P\prod_{\langle v,p\rangle} CZ_{vp} \ket{+}_{PV}
\end{align}
is the $+1$ eigenstate of the stabilizers
\begin{align}
    A_p &= {\color{blue}{x_p}}  \times \raisebox{-0.5\height}{\includegraphics[scale=0.9]{ApCC.pdf}},
    &B_p &=  \raisebox{-0.5\height}{\includegraphics[scale=0.9]{BpCC.pdf}}
\end{align}
For convenience, let us denote the six vertices surrounding the plaquettes as $1,\ldots,6$. Then,
\begin{align}
    A_p &= {\color{blue}{x_p}}  \times \prod_{n=1}^6 Z_n,
    &B_p &=  \prod_{n=1}^6 X_n & A_p B_p = - {\color{blue}{x_p}}  \times \prod_{n=1}^6 Y_n
\end{align}
Here, we have also included the product stabilizer to point out that the state inherently has a $\ZZ_2$ symmetry that swaps $A_p$ and $A_pB_p$ \textit{independent} of the measurement outcome. This symmetry is realized by acting with $\frac{Y+ Z}{\sqrt 2}$ on one sublattice and $\frac{Y- Z}{\sqrt 2}$ on the other sublattice. That is, on the six sites, it acts as $\prod_{n=1}^6 \frac{Y +(-1)^n Z}{\sqrt 2}$. (this sublattice structure is essential to obtain the minus sign in  $A_pB_p$).

In order to measure the Gauss law for this symmetry, it is helpful to perform a basis transformation to turn the symmetry $\prod_v \frac{Y\pm Z}{\sqrt 2}$ (where $\pm$ denotes the sublattice structure) into $\prod_v X_v$. This is accomplished by the second layer of the protocol: $ \prod_v e^{\pm \frac{\pi i}{8} Z_v} H_v$
 After the transformation, the state is given by stabilizers
\begin{align}
   \tilde  A_p &= {\color{blue}{x_p}}  \times \prod_{n=1}^6 \frac{X_n+(-1)^n Y_n}{\sqrt{2}}
    &\tilde B_p &=  \prod_{n=1}^6 Z_n,
    &\tilde  A_p\tilde B_p = {\color{blue}{x_p}}  \times \prod_{n=1}^6 \frac{X_n-(-1)^n Y_n}{\sqrt{2}}.
\end{align}
Indeed, $\prod_v X_v$ swaps $\tilde  A_p$ and $\tilde  A_p\tilde B_p$ while leaving $\tilde B_p$ invariant as desired.

Finally, in the last step we measure the Gauss law for this $\ZZ_2$ symmetry $X_v \prod_{e \supset v} Z_e$ on all vertices.

This can be done by initializing qubits on all edges in the $\ket{+}$ state, applying Controlled-$Z$ connecting vertices to all the nearest edges and measuring all the vertices in the X basis.

The new edges introduced are stabilized by $X_e$, and after applying $\prod_{v,e}CZ_{ve}$, the stabilizers are given by $\tilde  A_p C_p$ and $\tilde B_p$ for each plaquette where
\begin{align}
    C_p = \raisebox{-0.5\height}{\includegraphics[scale=0.9]{Cp.pdf}} \label{eq:Cp}
\end{align}
and $D_e = Z_v X_e Z_{v'}$ for each edge, where $v$ and $v'$ are the vertices at the end points of $e$. Now, to perform the measurement in the $X$ basis on all vertices, we need to find combinations of stabilizers that commute with the measurement. First, we note the following combinations do not involve vertex terms
\begin{align}
    B_p D_{12} D_{34} D_{56} &= X_{12} X_{34} X_{56} & B_p D_{23} D_{45} D_{61} &= X_{23} X_{45} X_{61}.
\end{align}
and therefore survives the measurement. Next, we consider the symmetric combination 
\begin{align}
    \tilde A_p C_p \frac{1+ \tilde B_p}{2} =  {\color{blue}{x_p}} \times \prod_{n=1}^6 X_n \times C_p \times  \left[\prod_{n=1}^6 \frac{1+(-1)^n iZ_n}{\sqrt{2}} + \prod_{n=1}^6 \frac{1-(-1)^n iZ_n}{\sqrt{2}} \right]
\end{align}
Expanding the bracket we find
\begin{align}
    \tilde A_p C_p \frac{1+ \tilde B_p}{2} =  {\color{blue}{x_p}} \times \prod_{n=1}^6 X_n \times C_p \times  \frac{1+ \tilde B_p}{2} \prod_{n=1}^6 \frac{1+Z_{n-1}Z_n + Z_{n}Z_{n+1} -  Z_{n-1}Z_{n+1}}{2}
\end{align}
Since the state satisfies $\tilde B_p =1$, it therefore also has eigenvalue +1 under the ``stabilizer"
\begin{align}
    {\color{blue}{x_p}} \times  \prod_{n=1}^6 X_n \times C_p \times  \prod_{n=1}^6 \frac{1+Z_{n-1}Z_n + Z_{n}Z_{n+1} -  Z_{n-1}Z_{n+1}}{2}
\end{align}
Note that these ``stabilizers" no longer commute amongst themselves. Now, using the fact that the state satisfies $Z_{n} X_{n,n+1} Z_{n+1}= 1$, we can replace $Z_{n}Z_{n+1}$ by$X_{n,n+1}$. This results in 
\begin{align}
    {\color{blue}{x_p}} \times  \prod_{n=1}^6 X_n \times C_p \times  \prod_{n=1}^6 \frac{1+X_{n-1,n} + X_{n,n+1} -  X_{n-1,n}X_{n,n+1}}{2}
\end{align}
This ``stabilizer" now commutes with the measurement on all vertices. With measurement outcomes $X_n = {\color{blue}x_n} = \pm 1$. To conclude, the final ``stabilizers" for each plaquette are
\begin{align}
     {\color{blue}{x_p}} \times \prod_{n=1}^6 {\color{red}x_n} \times C_p \times  \prod_{n=1}^6 \frac{1+X_{n-1,n} + X_{n,n+1} -  X_{n-1,n}X_{n,n+1}}{2},&&  X_{12} X_{34} X_{56},&& X_{23}X_{45} X_{61}.
\end{align}
Finally, performing Hadamard on all edges and using the fact that $CZ_{ij} = \frac{1+Z_i+Z_j-Z_iZ_j}{2}$, we recover the ``stabilizers" in Eq.~\eqref{eq:D4stabilizers}.

\section{Implementation on Sycamore}\label{app:Sycamore}

\begin{figure*}[h!]
    \centering
    \includegraphics[scale=0.6]{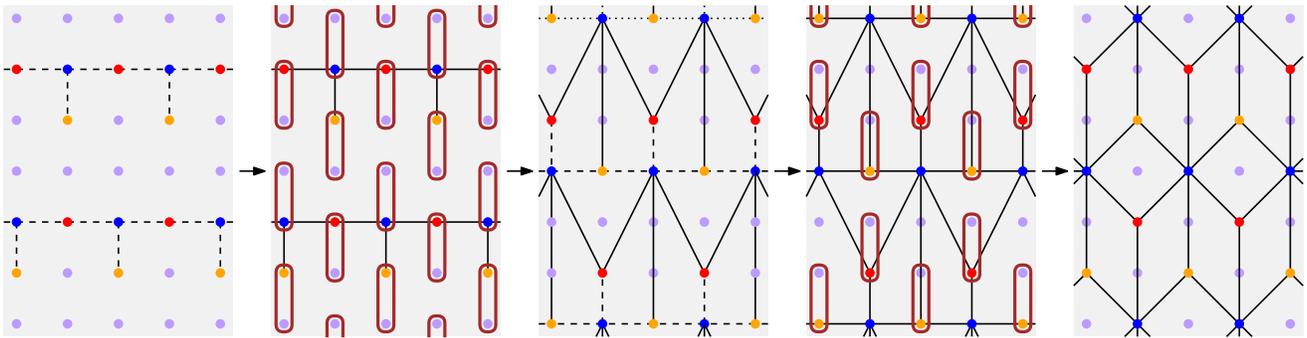}
    \caption{Preparation of the dice lattice cluster state in Sycamore using SWAP gates. Purple qubits only participate in the swapping procedure, and are not part of the cluster state.}
    \label{fig:diceprep}
\end{figure*}

First, let us count the depth of the 2-body gates required on the ideal lattice in Fig.~3a of the main text. The dice lattice cluster state can be prepared in depth 6, while the heavy-hex lattice cluster state can be prepared in depth 3. This gives a total 2-body depth count of 9.

Next, we discuss the details of implementation on the a quantum processor with connectivity of the square lattice, such as Google's Sycamore quantum chip. The first step in our protocol $\prod_{\inner{p,v}} CZ_{pv}$ requires preparing a cluster state on the dice lattice. This can be achieved by the help of SWAP gates. As seen in Fig.~\ref{fig:diceprep}, the four steps corresponds to steps 1,2,3 and 5 in Fig.~3b, and indeed produces the cluster state.

Next, we note that the single site rotation $e^{\pm \frac{\pi i}{4} Z_v}H_v$ can be pulled back through the final layer of SWAP gates, so that it acts on the corresponding sites \textit{before} the swap. This results in in step 4 of Fig.~3b. Lastly, $\prod_{\inner{v,e}}CZ_{ve}$ which forms the heavy-hex lattice is implemented in step 6.

To count the number of gates used, the $CZ$ gates in steps 1 and 3 can each be implemented in depth-3. Therefore, the 2-body gate depth count for the six steps combined is is 3+1+3+0+1+3=11.

More practically, we should count the 2-body gate depth using the innate gates of the Sycamore processor. In particular, the SWAP gate can be decomposed in to three $CZ$ gates interspersed by Hadamard gates. Conveniently, one of the $CZ$ gates from the SWAP in step $5$ exactly cancels one of the $CZ$ layers in step 6. Thus, the innate 2-body depth count is 3+3+3+0+2+2=13.

\section{Single-round preparation of topological orders with Lagrangian subgroup}

We give a formal argument that any non-Abelian topological order in two spatial dimensions which admits a Lagrangian subgroup can be prepared using a single round of measurements.

To recall, a Lagrangian subgroup $A$ is a subset of Abelian anyons that are closed under fusion, have trivial self and mutual statistics, and that every other anyon braids non-trivially with at least one of the anyons in the subgroup\cite{KapustinSaulina11,Levin13}. Note that the full set of anyons describing the theory does not need to be Abelian\footnote{In fact, we use the word \textit{subgroup} in contrast to the more general \textit{subalgebra} precisely because we restrict $A$ to only contain Abelian anyons.}. Before moving forward, we remark that $A$ can serve two purposes in this discussion: it can be a set that contains anyons, or can also function as an abstract group.

Given a topological order and a Lagrangian subgroup $A$, one can ``condense" \cite{burnell_anyon_2018} all the anyons in $A$. To do this, we introduce an auxiliary system with global symmetry given by the group $A$. The system has charges that transform under irreps of the global symmetry $a_\text{phys}$. Note that these charges are physical, unlike the anyons $a$ which are gauge charges of an unphysical gauge group. Next, one performs a condensation for all bound states $a \times a_\text{phys}^{-1}$. This identifies $a \sim  a_\text{phys}$ in the ground state of the condensed phase. The symmetry of the system is still $A$. However, the remaining anyons are confined, since they braid non-trivially with the anyons that are condensed. These confined anyons now serve as defects of the symmetry $A$. Since the resulting phase no longer has any anyons, it is therefore a (bosonic) Symmetry-Protected Topological (SPT) phase with global symmetry $A$. Let us call this state $\ket{\psi_{\SPT}}$ This process is also known as gauging the 1-form symmetry for all anyon lines in $A$\cite{KapustinSaulina11,Kaidi2021higher}\footnote{We remark that if one wishes, this condensation process can be implemented physically without tuning through a phase transition using measurements \cite{measureSPT}. The auxiliary degrees of freedom serve as ancillas for which the hopping that promotes the condensation can be measured.}

Now, we provide a protocol to prepare such a topological order. We start from a trivial product state with symmetry group $A$. It is known that any SPT phase in two spatial dimensions can be prepared (by temporarily breaking the symmetry) with a finite-depth local unitary\cite{chen_symmetry_2013}\footnote{up to possibly the $E_8$ phase, which ultimately decouples from the desired topological order}. Therefore, after preparing $\ket{\psi_\SPT}$, we measure the symmetry charges of $A$ by coupling the charges to ancillas so that we can measure its Gauss law. Note that this is nothing but the protocol to implement the Kramers-Wannier transformation in Ref. \onlinecite{measureSPT}. After the measurement, the charges of $\ket{\psi_\SPT}$ are promoted to gauge charges, and therefore realizes the anyons in $A$, and the symmetry fluxes are promoted to deconfined gauge fluxes, restoring the remaining anyons in the theory. In other words, our measurement has reversed the condensation by gauging the global symmetry $A$. To summarize, we have used finite-depth local unitaries and one round of measurement to prepare a state in the desired phase without feedforward or postselection. Note that if one moreover wants to prepare exactly the ground state of the phase, a single round of feedforward gates suffices to pair up all the anyons in $A$ that result from the measurement.

The condition of a Lagrangian subgroup can actually be relaxed if we allow physical fermions as resources. In particular, the subgroup $A^f$ can now contain anyons that have fermionic self-statistics, a fermionic Lagrangian algebra\cite{Bultinck20}. In this case, one performs ``fermion condensation"\cite{AasenLakeWalker2019}. For any fermionic anyon in $A^f$ the bound state that one condenses is now $f \times f_\text{phys}^{-1}$. This gives a fermionic SPT state  $\ket{\psi^f_{\SPT}}$  with global symmetry $A^f$, which contains fermion parity as a subgroup.

Similarly, starting with a trivial product state with fermionic symmetry $A^f$ it is possible to prepare the SRE state $\ket{\psi^f_\SPT}$ using finite-depth local unitaries. Then, one measures the Gauss law for this symmetry. For group elements that corresponds to anyons with fermionic statistics, measuring the Gauss law of the fermion amounts to performing the two-dimensional Jordan-Wigner transformation (bosonization)\cite{ChenKapustinRadicevic2018}, which can be performed using measurements\cite{measureSPT}.

\subsection{Example: $D_4$ TO revisited}\label{app:D4TQD}
To give a concrete example, let us consider the quantum double for $D_4$. The Lagrangian subgroup is given by the sign representations along with the conjugacy class of the center. These anyons form a group $A=\ZZ_2^3$. By performing condensation, one arrives at an invertible state with symmetry $\ZZ_2^3$. In fact, this state is a Symmetry-Protected Topological state, and can be deformed (while preserving the symmetry) to following hypergraph state \cite{Yoshida2016}
\begin{align}
   \ket{\psi_\SPT^{D_4}}= \prod_{\langle p_1p_2p_3\rangle } CCZ_{p_1p_2p_3} \ket{+}_P
   \label{eq:hypergraph}
\end{align}
where $\langle p_1p_2p_3\rangle$ denotes three plaquettes that share a common vertex\footnote{This SPT corresponds to the so-called type-III cocycle $a_1a_2a_3$.}. To describe the $\ZZ_2^3$ symmetry, we first note the plaquettes are three-colorable (say, red green and blue), such that no two adjacent plaquettes have the same color. Then each $\ZZ_2$ symmetry acts as spin flips on a plaquette of a fixed color.

To prepare the $D_4$ TO, we thus first prepare the above hypergraph state using $CCZ$. Then, we gauge the $\ZZ_2^3$ symmetry by measuring the Gauss law $\displaystyle X_p \prod_{e \in \raisebox{-0.4\height}{ \includegraphics[scale=0.25]{eforZ23.pdf}}}Z_e$, where $e$ are the six edges radiating out of each plaquette. This can be done via
\begin{align}
 \ket{D_4}_E =  \bra{+}_{P}\prod_{p, e \in\raisebox{-0.4\height}{ \includegraphics[scale=0.25]{eforZ23.pdf}} } CZ_{pe}   \ket{+}_{E} \ket{\psi_{\SPT}^{D_4}}
\end{align}
It is shown in Ref. \onlinecite{Yoshida2016} that the resulting state (after applying Hadamard on all edges) has exactly the same ``stabilizers" we derived in Eq.~\eqref{eq:D4stabilizers}. This corresponds to the fact that the $D_4$ TO can be regarded as a $\ZZ_2^3$ twisted quantum double\cite{DijkgraafPasquierRoche91TQD,Propitius95,HuWanWu13TQD}.

\subsection{Example: $Q_8$ TO}
As a second example, we consider the quantum double for the quaternion group $Q_8$. The Lagrangian subgroup also consists of sign representations along with the conjugacy class of the center, and forms the group $A=\ZZ_2^3$. After condensation, we arrive at a different SPT state, corresponding to the following hypergraph state
\begin{align}
   \ket{\psi_\SPT^{Q_8}}=\prod_{\langle p_1^Cp_2^Cp_3^C\rangle }  CCZ_{p_1^Cp_2^Cp_3^C} \prod_{\langle p_1p_2p_3\rangle } CCZ_{p_1p_2p_3} \ket{+}_P
   \label{eq:hypergraphQ8}
\end{align}
where $\langle p_1^Cp_2^Cp_3^C\rangle$ consists of three plaquettes of the same color connected by edges in a triangle shape\footnote{This SPT can be described by a combination of type-I and type-III cocycles $a_1^3 + a_2^3 + a_3^3 + a_1a_2a_3$ \cite{Propitius95}.}.

Thus, the $Q_8$ TO can be prepared as
\begin{align}
 \ket{Q_8}_E =  \bra{+}_{P}\prod_{p, e \in\raisebox{-0.4\height}{ \includegraphics[scale=0.25]{eforZ23.pdf}} } CZ_{pe}   \ket{+}_{E} \ket{\psi_{\SPT}^{Q_8}}
\end{align}

and the ``stabilizers" of this state after Hadamard is given by 
\begin{align}
    \bs A_p &= \raisebox{-0.5\height}{\includegraphics[scale=0.9]{ApQ81.pdf}} \times  \raisebox{-0.5\height}{\includegraphics[scale=0.9]{ApQ82.pdf}}, 
    & \bs B^{(1)}_p &= \raisebox{-0.5\height}{\includegraphics[scale=0.9]{Bp1D4.pdf}},
    & \bs B^{(2)}_p &= \raisebox{-0.5\height}{\includegraphics[scale=0.9]{Bp2D4.pdf}}.
    \label{eq:Q8stabilizers}
\end{align}

\subsection{Example: Double Ising TO}

As a final example, we discuss how to prepare the Doubled Ising TO, which is obtained by stacking the Ising TO consisting of anyons $\{1, \sigma, \epsilon \}$ with its time-reversed partner $\{1, \bar \sigma, \bar \epsilon \}$. Since $\epsilon$ and $\bar \epsilon$ are fermions, a Lagrangian subgroup  does not exist. Nevertheless, it does have a fermionic Lagrangian subgroup $A^f = \{1,\epsilon, \bar \epsilon, \epsilon \bar \epsilon\}$. Condensing the fermionic Lagrangian subgroup results in an SPT state with $\ZZ_2 \times \ZZ_2^F$ symmetry. The precise wavefunction of this SPT state can be found in Ref. \onlinecite{TarantinoFidkowski16}, and since it is short-range entangled, it can be prepared with a finite-depth quantum circuit.

\end{document}